\documentclass[%
 aip,
 amsmath,amssymb,
 reprint,%
]{revtex4-1}

\usepackage{graphicx}
\usepackage{dcolumn}
\usepackage{bm}

\usepackage[utf8]{inputenc}
\usepackage[T1]{fontenc}
\usepackage{mathptmx}
\usepackage{color}
\usepackage[normalem]{ulem}

\newcommand{\vect}[1]{\mathbf{#1}}

\begin{document}


\title{Plenoptic x-ray microscopy}

\affiliation{Institute of Physics, Jagiellonian University, \L{}ojasiewicza 11, 30-348 Krak\'ow, Poland
}%

\author{Katarzyna M. Sowa}
\author{Marcin P. Kujda}%
\author{Pawe\l{} Korecki}
\email{pawel.korecki@uj.edu.pl.}

\date{\today}

\begin{abstract}
Plenoptic cameras use arrays of micro-lenses to capture multiple views of the same scene in a single compound image. They enable refocusing on different planes and depth estimation. However, until now, all types of plenoptic computational imaging have been limited to visible light.
We demonstrate an x-ray plenoptic microscope that uses a concentrating micro-capillary array instead of a micro-lens array and can simultaneously acquire from one hundred to one thousand x-ray projections of imaged volumes that are located in the focal spot region of the micro-capillary array. Hence, tomographic slices at various depths near the focal plane can be reconstructed in a way similar to tomosynthesis, but from a single x-ray exposure. The microscope enables depth-resolved imaging of small subvolumes in large samples and can be used for imaging of weakly absorbing artificial and biological objects by means of propagation phase-contrast.
\end{abstract}

\maketitle

Due to their large penetration depth and short wavelength, x-rays can be used to look deep inside  matter at high spatial resolution. However, a single x-ray radiograph is a 2D projection of a 3D object. Therefore, some radiographs are too complex for analysis due to the overlap of shadows of features that are located at different depths of the sample. Three-dimensional x-ray imaging with isotropic spatial resolution is possible  using computed tomography (CT) \cite{Kak,Sakdinawat2010,Shearer2016}.  A CT scan is realized by rotation of the sample or the x-ray source and the detector and via the collection of hundreds x-ray projections for various viewing angles in the full angular range of 180$^\circ$ or 360$^\circ$. It is also possible to focus on slices of interest via limited-angle tomography. The reduction in the angular range enables the acceleration of  imaging and the reduction of the radiation dose and yields high in-plane resolution at the expense of limited depth resolution \cite{Wu2003}. For example, x-ray tomosynthesis is an important tool in medicine \cite{Dobbins2009}, and x-ray laminography is implemented in the semiconductor industry \cite{Gondrom1999}. The collection of a limited set of projections still requires angle-by-angle acquisition. Further acceleration of acquisition can be realized by using multi-source or scanning x-ray tubes, which are flashed sequentially \cite{Neculaes2014}.


In recent works, a close relation between limited-angle cone-beam tomography and the so-called plenoptic or light-field camera has been identified \cite{Vigano2018,Vigano2019}. A plenoptic camera captures both the intensity and direction of light-rays and uses arrays of micro-lenses to collect multiple views of the same scene in a single image \cite{Lippmann1908,Adelson1992,Ng2005,Georgiev2010,Lam2015}. Optical microscopy \cite{Levoy2009} and correlation imaging \cite{Angelo2016} with plenoptic cameras is also possible. Extension of plenoptic imaging to x-ray range is cumbersome and has not yet been presented. Despite the huge progress in x-ray optics development \cite{Sakdinawat2010,Snigirev1996}, due to the very small and negative refractive index decrement \cite{paganin}, 2D arrays of x-ray lenses are difficult to fabricate. However, the negative refractive index decrement enables the highly efficient guidance of  x-rays inside glass micro-capillaries \cite{Bilderback1994,Vincze1998} or capillary arrays \cite{Chapman1991} via total external reflections.

Polycapillary devices consist of hundreds of thousands of such thermally shaped microcapillaries and can collimate or focus x-rays \cite{KumakhovPhysicsReports1990,MacDonald2017}. In our group, it was demonstrated \cite{Dabrowski2013APL} that polycapillary optics can be used for x-ray microlaminography using the coded aperture principle \cite{Fenimore1978,Nadjmi1980,DabrowskiOE2013}. However, depth resolution was achieved by scanning the sample in both the lateral and axial directions and the main objective of single exposure 3D imaging has not been realized. Moreover, coded aperture approach has strong disadvantages, and the weak signal-to-noise ratio made the imaging of weak features via phase-contrast impractical \cite{Sowa2017}. More recently, it was demonstrated that individual capillaries \cite{Sowa2018} (or even defects \cite{Korecki2017}) inside a polycapillary optics can be used to generate an array of secondary quasi-point x-ray sources and to obtain replicated projections of planar objects that are placed at the focal plane in the so-called multipoint projection geometry. Here, by using multipoint projection we realize plenoptic x-ray microscopy, which enables single exposure depth-resolved imaging of sample subvolumes in the focal spot region of a polycapillary optics.

\begin{figure}[tbh]
\centering\includegraphics[width=8cm]{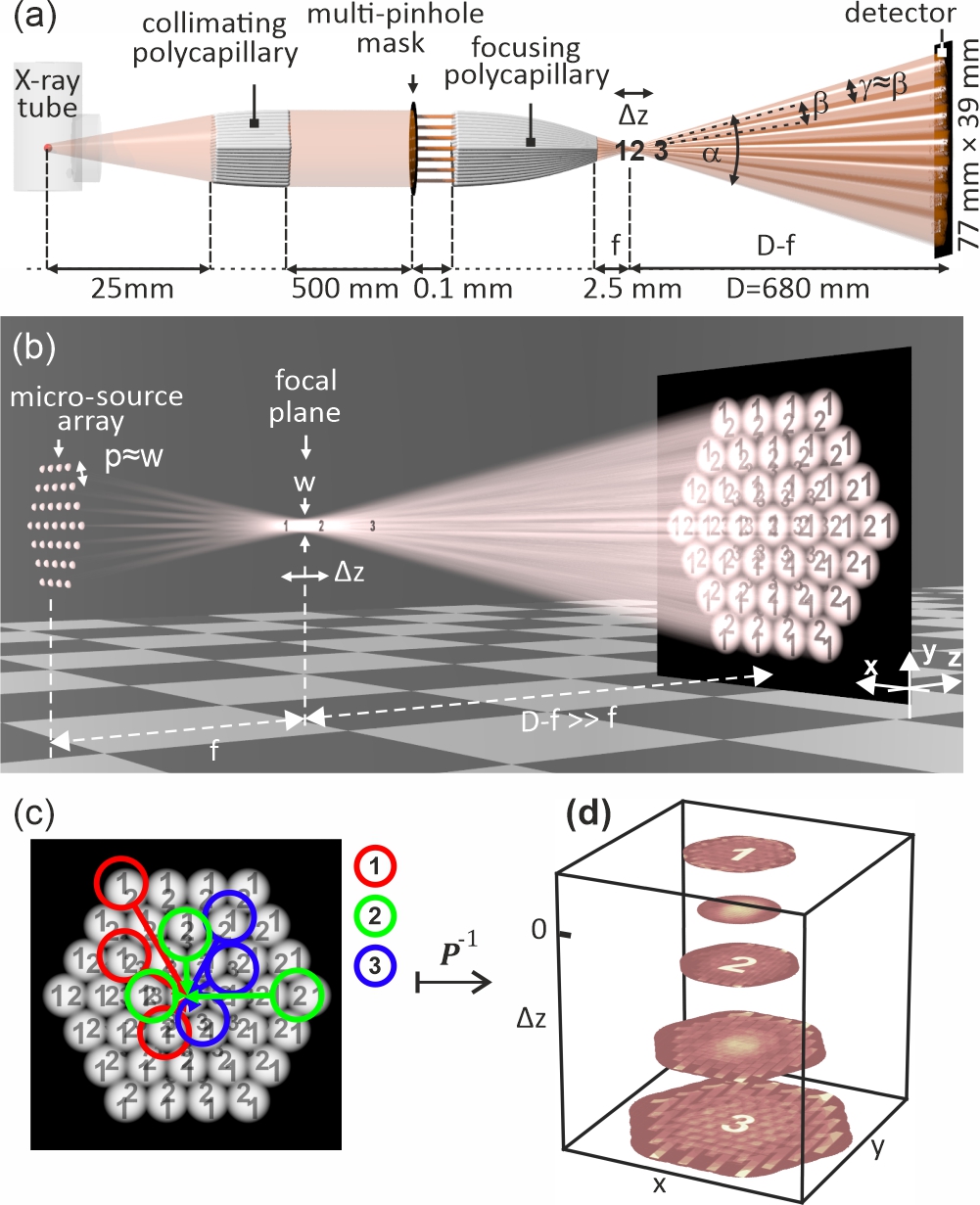}
\caption{Principle of plenoptic x-ray microscopy. (a) A simplified diagram of the experimental setup. The key element is a polycapillary focusing optics, which generates an array of secondary x-ray microsources. A multi-pinhole mask located in the front of the focusing polycapillary optics selects a sparse distribution of capillaries according to condition $p\approx w$. (b) Geometrical ray-tracing of image formation. (c) The idea of plenoptic reconstruction - depth-dependent affine transformation that involves translation of sub-images and scaling operation. Translation vectors for different $\Delta z$ planes are shown with different colors. (d) The plenoptic reconstruction at various depths.}\label{fig:fig1}
\end{figure}

In the experimental setup [Fig. \ref{fig:fig1}(a)], a collimating polycapillary optics is attached to an x-ray tube and generates a    quasi-parallel x-ray beam. Here, we used an x-ray tube with a focal spot of $~40$~$\mu$m and a tungsten anode that generated a highly polychromatic spectrum (W L lines and bremsstrahlung) with a mean energy of $~\sim$~9 keV.
\begin{figure}[t]
\centering\includegraphics[width=8.5cm]{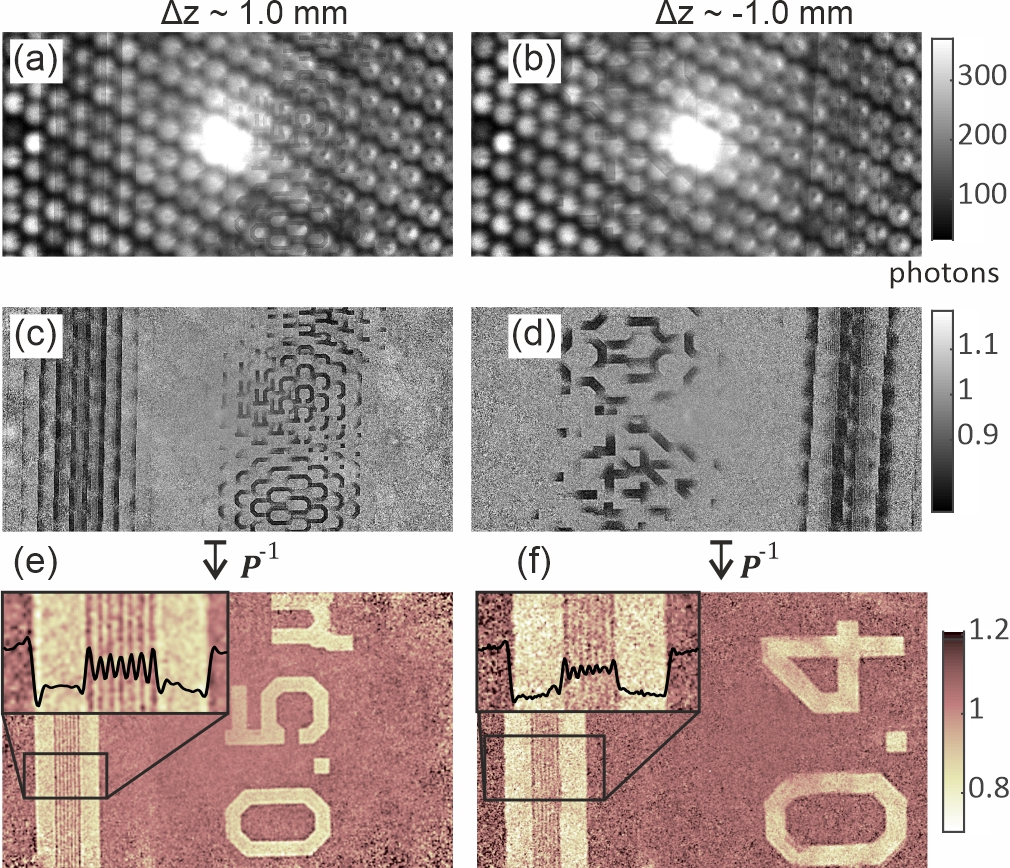}
\caption{Test of the x-ray plenoptic reconstruction for a planar object (JIMA-RT-02B resolution chart). The images on the left  (a,c,e) correspond to slits with 0.5~$\mu$m line widths that are located at $\Delta z\sim$1 mm (magnification of $M\sim$194),  and images on the right (b,d,f) to slits with 0.4~$\mu$m line widths that are located at $\Delta z\sim$-1 mm (magnification of $M\sim 453$). (a,b) Raw data. (c,d) The data after normalization.  (e,f) The plenoptic reconstruction. }\label{fig:fig2}
\end{figure}
\begin{figure*}[t]
\centering\includegraphics[width=17cm]{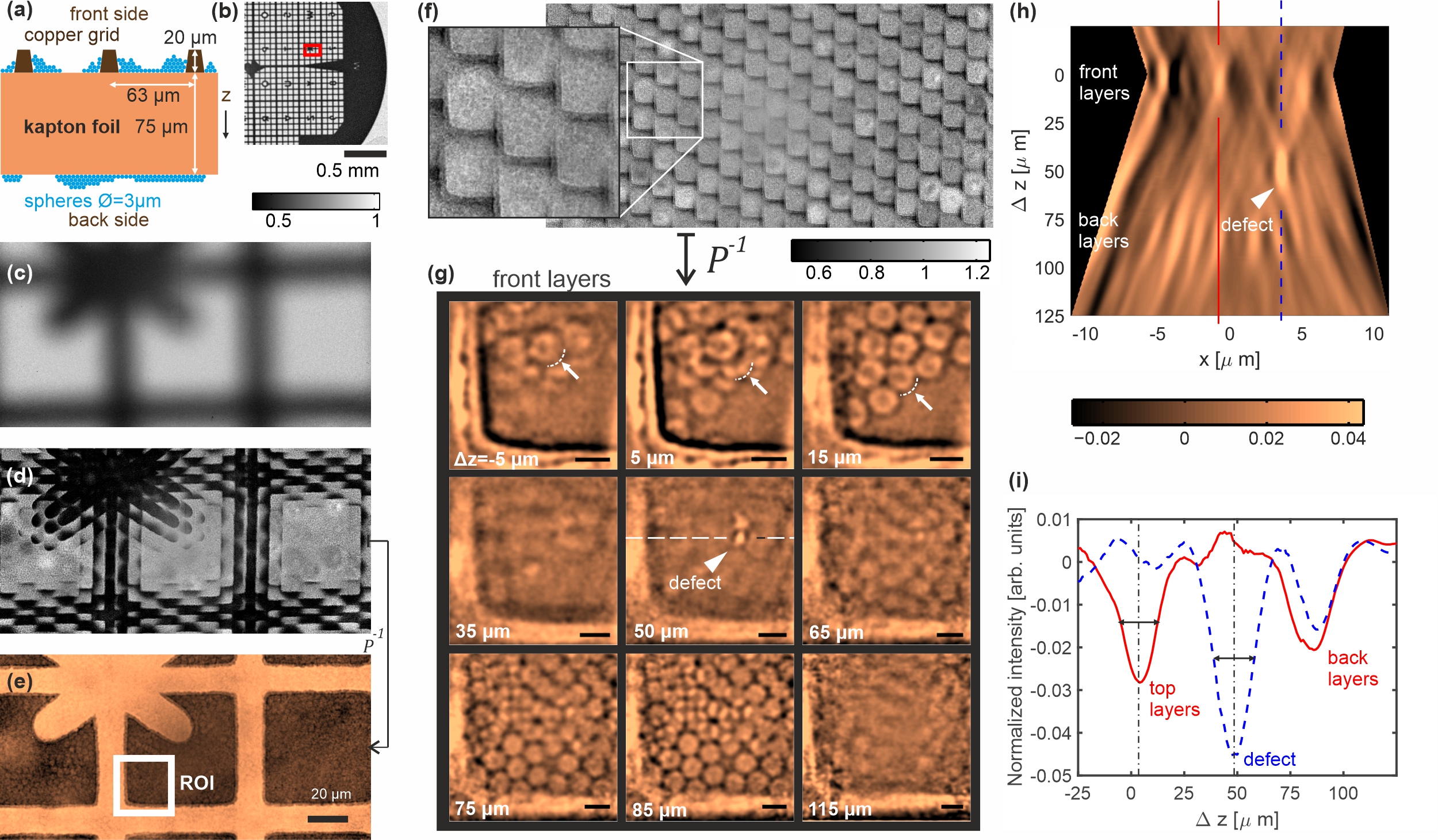}
\caption{Reconstruction of tomographic slices from x-ray plenoptic images.  (a) A sketch of the sample. (b,c) Conventional x-ray projections of the sample without the multi-pinhole mask. (d) An x-ray image that was recorded with the multi-pinhole mask for $\Delta z \approx$1.5 mm, at a geometrical magnification $M\approx 170$.  (e) The plenoptic reconstruction for $\Delta z\gg 0$ represents a high-resolution projection of the sample. (f) A plenoptic image for $\Delta z \sim$0~$\mu$m  ($M\sim 270$) for the region-of-interest (ROI) marked in (e). This image is composed of multiple  projections of nearly the same region of the sample that differ in terms of the viewing
angle. (g) Plenoptic reconstruction of lateral tomographic slices. Each image shows a reconstructed tomographic slice at the $\Delta z$ value that is indicated in the text label. Arrows that point to dashed arcs show spheres in successive layers. The triangle marks a defect inside the kapton. Scale bars: 5~$\mu$m. (h) $z$-axis plenoptic reconstruction at the position of the defect. (i) Line profiles of the slice from (h). Double arrows extend by $\pm$10~$\mu$m. (Multimedia view)
} \label{fig:fig3}
\end{figure*}
A multi-pinhole mask selects a sparse array of micro-beams, which are focused with a second glass polycapillary element and form a structured x-ray field i.e. a short wavelength counterpart of a light-field\cite{Lam2015}.
In the experiment, this x-ray field  was captured with a direct-detection hybrid pixel counting detector (Dectris Eiger2 R 500k with pixel pitch of 75~$\mu$m). While the concept of a plenoptic camera can be generalized to Fourier optics \cite{Lam2015}, the basic principle relies on geometrical optics, which is wavelength-independent. Therefore, to illustrate the principle of an x-ray plenoptic microscope, we used a standard ray-tracing software (POV-Ray). An array of secondary spotlight sources that are generated at tips of capillaries illuminate a common focal point [Fig. \ref{fig:fig1}(b)]. A 3D object is modelled as a stack of semi-transparent numbers (1, 2 and 3).

In the multi-point projection geometry, assuming a discrete array of equally intense point-sources at positions $\vect{r}_i=(x_i,y_i )$, the image in the detector at position $\mathbf{R}$ can be expressed as:
\begin{equation}\label{eq:eq1}
I(\vect{R})\approx \frac{1}{D^2} \sum_{i=1}^{N} T\left (\frac{\vect{R}-\vect{r}_i}{D}d+\vect{r}_i \right )G\left(\vect{R}-m\vect{r}_i\right),
\end{equation}
where $m=(f-D)/f$, $D$ is the optics-to-detector distance, $f$ is focal length of the optics, $T$ is the transmission of a planar object that is located at a plane $d=f+\Delta z$, and $G$ describes the angular distribution of radiation that is generated by the point sources and has an angular width of $\gamma\sim 2\theta_{c}$, where $\theta_c$ is the critical angle for the total external reflection. A more exact version of Eq.~(\ref{eq:eq1}) was discussed in Ref.~\onlinecite{Korecki2015}, and the description of the propagation phase-contrast signal was provided in Ref.~\onlinecite{Sowa2018}.

If the spacing $p$ between the micro-sources is approximately equal to beam width at the focal plane $w$ here, $w\approx  12$~$\mu$m), all beams overlap at the focal plane but there is no overlap of the beams at the detector. In such a case, the image in the detector is equivalent to a set of truncated limited angle tomographic projections that are acquired in a single image. Reconstruction of tomographic slices at specified depth $\Delta z$ is possible via plenoptic reconstruction $(P^{-1})$, which is based on a depth dependent affine transformation and is similar to refocusing procedures that are implemented in visible-light plenoptic cameras \cite{Ng2005}. First, for each reconstruction slice at a plane $d$, an image that was generated by the $i$-th source was masked by a Gaussian function that is centered on $m\vect{r}_i$. Second, the experimental pattern was transformed or back-projected from the detector plane to the plane at $z=d$ using an affine transform with the \verb"imwarp" function in MATLAB's Image Processing Toolbox. The affine transformation was defined by the following augmented matrix:
\begin{equation}
\begin{bmatrix}
D/d & 0 & 0 \\
0 & D/d & 0 \\
(1-D/d)x_i & (1-D/d)y_i & 1
\end{bmatrix}^{-1}
 \label{eq:eq2}
\end{equation}
The transformations for all $i=1..N$ beams were summed. Finally, a similar operation was performed for an image that was filled with ones, which was used for normalization. An optional  bandpass filter could be applied to the final image for noise removal or background suppression. A schematic illustration of the transform is presented in Fig.~\ref{fig:fig1}(c,d).
\begin{figure*}[t]
\centering\includegraphics[width=16cm]{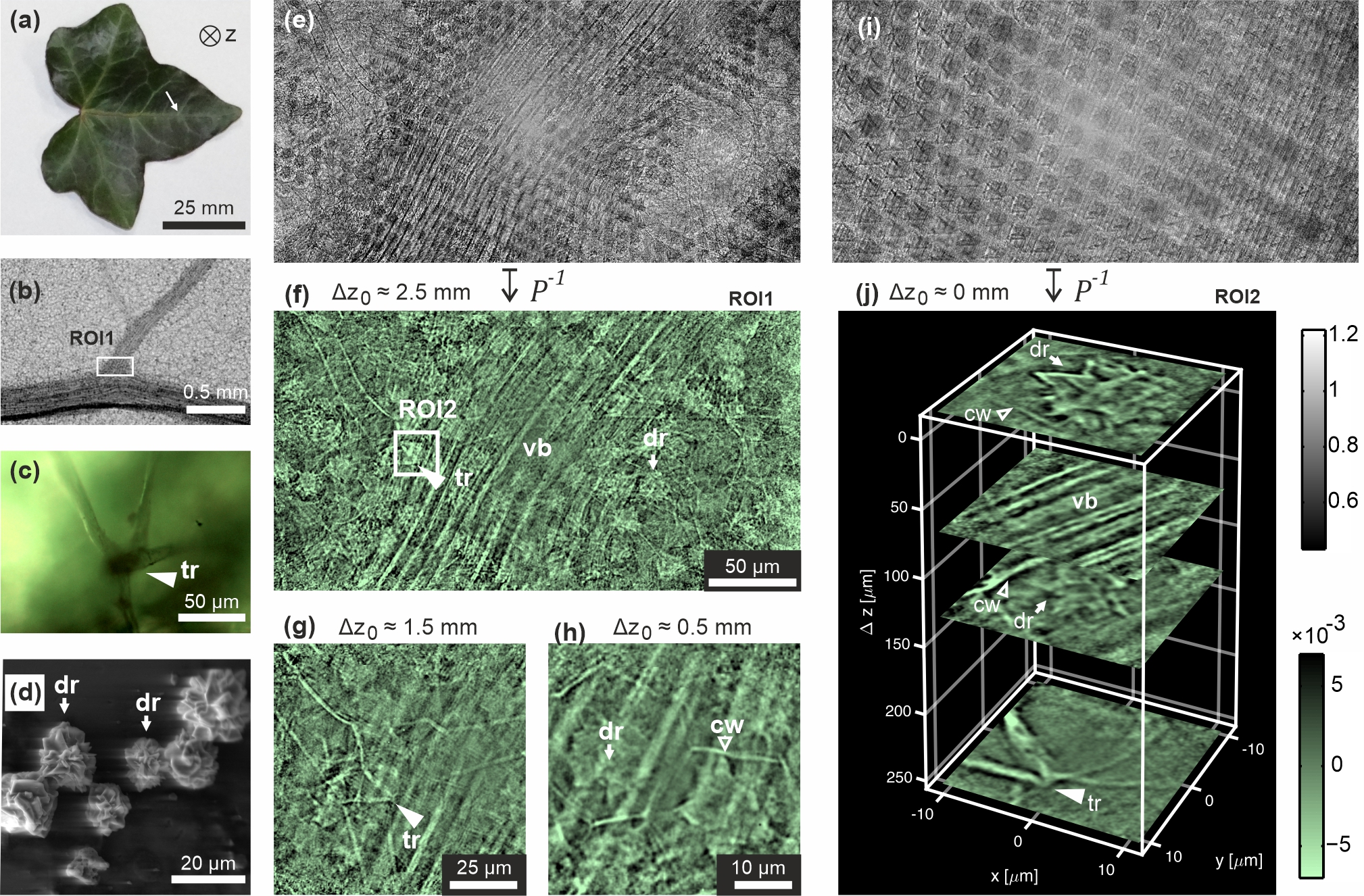}
\caption{Plenoptic x-ray microscopy of a biological sample - an ivy leaf. (a) A photograph of the leaf. The arrow points to the imaged region. (b) A conventional x-ray radiograph. (c) A trichome at  the bottom side - optical microscopy image. (d) A SEM image of druses that are isolated from an ivy leaf. (e,f) A plenoptic x-ray image and reconstruction for ROI1. (g,h)  Examples of plenoptic reconstructions for various positions of the sample. (i,j) A plenoptic image and central region of plenoptic reconstruction for ROI2. Color bars correspond to plenoptic data. Labels: tr - trichome, dr - druse, vb - vascular bundle, cw - cell wall, and $\Delta z_0$ - the position of the front surface of the sample relative to the focal plane. (Multimedia view) }\label{fig:fig4}
\end{figure*}

To test the accuracy and resolution of the plenoptic x-ray reconstruction, x-ray plenoptic imaging was performed on a planar resolution chart (JIMA-RT-02B) as shown in Fig.~\ref{fig:fig2}. The JIMA chart consists of 1~$\mu$m thick tungsten lines with various widths that are deposited on a 60~$\mu$m silicon wafer. Images for 0.5~$\mu$m slits and  0.4~$\mu$m slits were recorded for different $\Delta z$ positions of the resolution chart relative to the focal plane  The 0.4~$\mu$m line widths are the finest features on the JIMA chart and can be resolved with our microscope. The lateral resolution is limited by the sizes of individual micro-sources (the capillary channel diameters). Note, that individual subimages are non-inverted for $\Delta z>0$ and inverted for $\Delta z<0$. This effect resembles the image formation in a standard plenoptic camera, which is based on micro-lens arrays\cite{Georgiev2010}.

3D imaging was tested on a specially prepared object (Fig.~\ref{fig:fig3}) {(Multimedia view). A copper finder grid was attached to a kapton foil and 3~$\mu$m diameter latex spheres were deposited on both sides of the sample. By changing the distance of the object from the focal plane, one can trade-off between images with a larger field-of-view and images with a shallower depth-of-field and a higher resolution. In Fig.~\ref{fig:fig3}(e), the depth-resolution or the effective slice thickness is large and the reconstructed image shows a projection over the whole sample thickness in the region near the X on the grid. The grainy image structure results from the latex spheres that are imaged and is due to propagation phase contrast \cite{Snigirev1995,Cloetens1996}, which is required  for the imaging of light objects via sensitivity to changes of the real part of the refractive index \cite{Shearer2016}.
A crucial prerequisite for measuring the phase contrast is a high degree of transverse spatial coherence of the incident beam, the  requirements for chromatic  coherence  are much less stringent \cite{Pogany1997,Gureyev2001}. In plenoptic x-ray microscopy the partial coherence ($\sim 1 \mu m$ at the sample plane) is achievied via polychromatic irradiation by a set of mutually incoherent but partially coherent secondary sources that are produced at tips of individual capillaries, with sizes of $\sim 0.7$~$\mu$m.

An important example of x-ray plenoptic image is presented in Fig.~\ref{fig:fig3}(f). The sample was placed near the focal plane (the position of the sample's front  surface $\Delta z_{0} \sim$~0~$\mu$m). The plenoptic image contains multiple projections of nearly the same region of the grid that differ in terms of the viewing angle. The angular range of the projections is $\sim6.5^{\circ}\times 3.25^{\circ}$. With the total flux in the beam of $\sim 10^5$~photons/s, this plenoptic image was acquired in approx. 7200 seconds. The total number of projections generated by the capillary optics is around 1000 (see supplementary material). For high-resolution imaging, the detector was placed far from the sample, which effectively reduced this number to $\sim 160$. Use of a larger area detector will enable to capture the whole beam and to increase the maximum viewing angle. Based on image from Fig.~\ref{fig:fig3}(f), we reconstructed slices of the object at various depths [Fig.~\ref{fig:fig3}(g)]. Spheres at the front and the back sides of the kapton foil can be easily distinguished. An isolated defect inside the kapton was also imaged. For spheres at the front side of the sample, the depth resolution is sufficient for sequencing successive layers of the 3~$\mu$m diameter latex spheres, as marked by arrows that point to dashed arcs. Depending on the sphere stacking, successive layers are separated by $\sim~2.45$~$\mu$m or 3~$\mu$m.

For a more precise determination of the depth resolution, Fig.~\ref{fig:fig3}(h) shows a plenoptic slice in the depth direction. The reconstructed region has a shape determined by the waist of the x-ray beam. As in any angle-limited tomographic geometry, the depth resolution is much lower than the lateral resolution. The highest resolution is obtained in the focal spot region, i.e, in the depth-region of $|\Delta z| \lesssim{w/\alpha}$ (here, approx. $\pm 60$~$\mu$m), where microbeams from all capillaries overlap.  In the focal spot region, the depth of field (effective slice half-width) is still smaller than 10~$\mu$m as shown in Fig.~\ref{fig:fig3}(h). The depth of field increases with $\Delta z$. Therefore, individual layers of spheres at the back side are not resolved. Moreover, for larger $\Delta z$, the limited depth-resolution is manifested as an elongation of all reconstructed features not along the z-axis (as for smaller $\Delta z$), but along the illumination direction of individual capillaries, as shown in Fig.~3(h).
Also note that the phase contrast improves the depth resolution. In limited angle tomography, the depth resolution increases with the lateral frequency of the signal \cite{Dabrowski2013APL,Grant1972}. The propagation-based phase-contrast signal is approximately proportional to the Laplacian of the phase, which enhances the high-frequency signal components and which leads to an edge enhancement effect \cite{paganin,Gureyev2009}.

To demonstrate that the plenoptic x-ray microscopy can be used to image details of complex biological samples at the cellular level we imaged an ivy leaf [\emph{Hedera helix}, Fig.~\ref{fig:fig4}] (Multimedia view). An easy adaptation of the field of view and depth of field is especially useful in such a case. A larger area projection of the leaf near a secondary vein shows the complex structure of this sample: vascular bundles, druses  (calcium oxalate crystal aggregates) \cite{Franceschi2005,Costa2009} and a trichome (leaf hair) \cite{Ackerfield2001} on the back side. Translation of the leaf towards the focal plane revealed additional details of the sample, e.g., the fine structure of the trichome [Fig.~\ref{fig:fig4}(g)] and  the positions of druses inside hexagonal cells [Fig.~\ref{fig:fig4}(h)]. At the highest depth resolution, a single plenoptic image enables one to perform virtual paradermal sectioning. Details of druses, bundle fibers and a small fragment of the trichome are resolved in the depth direction as shown in Fig.~\ref{fig:fig4}(j). The plenoptic image from Fig.~\ref{fig:fig4}(i) was acquired in approx. 3600 seconds, and the radiation dose was estimated to $<10$~kGy \cite{kirz1995}.

In conclusion, we extended plenoptic imaging to the hard x-ray range. Due to the penetrating nature of x-rays, it is suited not only for depth estimation, but enable 3D imaging of opaque objects and has a potential of much higher lateral resolution. Due to the limited angular apertures of polycapillary devices, the depth resolution will be always much worse than the lateral resolution. Therefore, plenoptic x-ray imaging is well suited for focusing at particular slices or regions of 3D volumes, in a way similar to zonography \cite{DANIELS1996} or tomosynthesis\cite{Vedantham2015}.  Future applications of plenoptic x-ray  microscopy will require a combination of the plenoptic reconstruction with  phase retrieval algorithms.  For 2D data recorded in multi-point projection geometry \cite{Sowa2018}, a phase retrieval based on Paganin filter \cite{Paganin2002} was applied. For 3D imaging more advanced algorithms could be required \cite{Thompson2019,Ming2019}. A more powerful x-ray source (a rotating anode \cite{Dehlinger2013,Brombal2019}, metal jet source \cite{Bauer2018} or a diamond hybrid anode x-ray tube \cite{Graf2018}) and a larger detector (up to 16M pixel) \cite{Casanas2016} could decrease the acquisition times by over an order of magnitude. Since the experimental setup is compatible with a parallel beam illumination [Fig.~\ref{fig:fig1}(a)], the use of synchrotron radiation would permit time-resolved single-shot studies \cite{Villanueva-Perez2018,Duarte2019}. Developments in energy-resolved hybrid pixel detectors could also enable hyperspectral 3D chemical imaging \cite{Egan2015}

See the supplementary material for experimental details, sample preparation and discussion of the resolution.

This work was financially supported by the Polish National Science Centre (grant no. DEC-2017/25/B/ST2/00152). K. M. S acknowledges support from Polish Ministry of Science and Higher Education (project no. 7150/E-338/M/2018). We thank B.R. Jany for performing the SEM imaging.


\begin{thebibliography}{54}%
\makeatletter
\providecommand \@ifxundefined [1]{%
 \@ifx{#1\undefined}
}%
\providecommand \@ifnum [1]{%
 \ifnum #1\expandafter \@firstoftwo
 \else \expandafter \@secondoftwo
 \fi
}%
\providecommand \@ifx [1]{%
 \ifx #1\expandafter \@firstoftwo
 \else \expandafter \@secondoftwo
 \fi
}%
\providecommand \natexlab [1]{#1}%
\providecommand \enquote  [1]{``#1''}%
\providecommand \bibnamefont  [1]{#1}%
\providecommand \bibfnamefont [1]{#1}%
\providecommand \citenamefont [1]{#1}%
\providecommand \href@noop [0]{\@secondoftwo}%
\providecommand \href [0]{\begingroup \@sanitize@url \@href}%
\providecommand \@href[1]{\@@startlink{#1}\@@href}%
\providecommand \@@href[1]{\endgroup#1\@@endlink}%
\providecommand \@sanitize@url [0]{\catcode `\\12\catcode `\$12\catcode
  `\&12\catcode `\#12\catcode `\^12\catcode `\_12\catcode `\%12\relax}%
\providecommand \@@startlink[1]{}%
\providecommand \@@endlink[0]{}%
\providecommand \url  [0]{\begingroup\@sanitize@url \@url }%
\providecommand \@url [1]{\endgroup\@href {#1}{\urlprefix }}%
\providecommand \urlprefix  [0]{URL }%
\providecommand \Eprint [0]{\href }%
\providecommand \doibase [0]{http://dx.doi.org/}%
\providecommand \selectlanguage [0]{\@gobble}%
\providecommand \bibinfo  [0]{\@secondoftwo}%
\providecommand \bibfield  [0]{\@secondoftwo}%
\providecommand \translation [1]{[#1]}%
\providecommand \BibitemOpen [0]{}%
\providecommand \bibitemStop [0]{}%
\providecommand \bibitemNoStop [0]{.\EOS\space}%
\providecommand \EOS [0]{\spacefactor3000\relax}%
\providecommand \BibitemShut  [1]{\csname bibitem#1\endcsname}%
\let\auto@bib@innerbib\@empty
\bibitem [{\citenamefont {Kak}\ and\ \citenamefont {Slaney}(2001)}]{Kak}%
  \BibitemOpen
  \bibfield  {author} {\bibinfo {author} {\bibfnamefont {A.~C.}\ \bibnamefont
  {Kak}}\ and\ \bibinfo {author} {\bibfnamefont {M.}~\bibnamefont {Slaney}},\
  }\href@noop {} {\emph {\bibinfo {title} {Principles of Computerized
  Tomographic Imaging}}}\ (\bibinfo  {publisher} {Society of Industrial and
  Applied Mathematics},\ \bibinfo {year} {2001})\BibitemShut {NoStop}%
\bibitem [{\citenamefont {Sakdinawat}\ and\ \citenamefont
  {Attwood}(2010)}]{Sakdinawat2010}%
  \BibitemOpen
  \bibfield  {author} {\bibinfo {author} {\bibfnamefont {A.}~\bibnamefont
  {Sakdinawat}}\ and\ \bibinfo {author} {\bibfnamefont {D.}~\bibnamefont
  {Attwood}},\ }\href@noop {} {\bibfield  {journal} {\bibinfo  {journal} {Nat.
  Photonics}\ }\textbf {\bibinfo {volume} {4}},\ \bibinfo {pages} {840}
  (\bibinfo {year} {2010})}\BibitemShut {NoStop}%
\bibitem [{\citenamefont {Shearer}\ \emph {et~al.}(2016)\citenamefont
  {Shearer}, \citenamefont {Bradley}, \citenamefont {Hidalgo-Bastida},
  \citenamefont {Sherratt},\ and\ \citenamefont {Cartmell}}]{Shearer2016}%
  \BibitemOpen
  \bibfield  {author} {\bibinfo {author} {\bibfnamefont {T.}~\bibnamefont
  {Shearer}}, \bibinfo {author} {\bibfnamefont {R.~S.}\ \bibnamefont
  {Bradley}}, \bibinfo {author} {\bibfnamefont {L.~A.}\ \bibnamefont
  {Hidalgo-Bastida}}, \bibinfo {author} {\bibfnamefont {M.~J.}\ \bibnamefont
  {Sherratt}}, \ and\ \bibinfo {author} {\bibfnamefont {S.~H.}\ \bibnamefont
  {Cartmell}},\ }\href {\doibase 10.1242/jcs.179077} {\bibfield  {journal}
  {\bibinfo  {journal} {Journal of Cell Science}\ }\textbf {\bibinfo {volume}
  {129}},\ \bibinfo {pages} {2483} (\bibinfo {year} {2016})}\BibitemShut
  {NoStop}%
\bibitem [{\citenamefont {Wu}\ \emph {et~al.}(2003)\citenamefont {Wu},
  \citenamefont {Stewart}, \citenamefont {Stanton}, \citenamefont {McCauley},
  \citenamefont {Phillips}, \citenamefont {Kopans}, \citenamefont {Moore},
  \citenamefont {Eberhard}, \citenamefont {Opsahl-Ong}, \citenamefont
  {Niklason},\ and\ \citenamefont {Williams}}]{Wu2003}%
  \BibitemOpen
  \bibfield  {author} {\bibinfo {author} {\bibfnamefont {T.}~\bibnamefont
  {Wu}}, \bibinfo {author} {\bibfnamefont {A.}~\bibnamefont {Stewart}},
  \bibinfo {author} {\bibfnamefont {M.}~\bibnamefont {Stanton}}, \bibinfo
  {author} {\bibfnamefont {T.}~\bibnamefont {McCauley}}, \bibinfo {author}
  {\bibfnamefont {W.}~\bibnamefont {Phillips}}, \bibinfo {author}
  {\bibfnamefont {D.~B.}\ \bibnamefont {Kopans}}, \bibinfo {author}
  {\bibfnamefont {R.~H.}\ \bibnamefont {Moore}}, \bibinfo {author}
  {\bibfnamefont {J.~W.}\ \bibnamefont {Eberhard}}, \bibinfo {author}
  {\bibfnamefont {B.}~\bibnamefont {Opsahl-Ong}}, \bibinfo {author}
  {\bibfnamefont {L.}~\bibnamefont {Niklason}}, \ and\ \bibinfo {author}
  {\bibfnamefont {M.~B.}\ \bibnamefont {Williams}},\ }\href {\doibase
  10.1118/1.1543934} {\bibfield  {journal} {\bibinfo  {journal} {Medical
  Physics}\ }\textbf {\bibinfo {volume} {30}},\ \bibinfo {pages} {365}
  (\bibinfo {year} {2003})}\BibitemShut {NoStop}%
\bibitem [{\citenamefont {Dobbins}(2009)}]{Dobbins2009}%
  \BibitemOpen
  \bibfield  {author} {\bibinfo {author} {\bibfnamefont {J.~T.}\ \bibnamefont
  {Dobbins}},\ }\href {\doibase 10.1118/1.3120285} {\bibfield  {journal}
  {\bibinfo  {journal} {Med. Phys.}\ }\textbf {\bibinfo {volume} {36}},\
  \bibinfo {pages} {1956} (\bibinfo {year} {2009})}\BibitemShut {NoStop}%
\bibitem [{\citenamefont {Gondrom}\ \emph {et~al.}(1999)\citenamefont
  {Gondrom}, \citenamefont {Zhou}, \citenamefont {Maisl}, \citenamefont
  {Reiter}, \citenamefont {Kr{\"o}ning},\ and\ \citenamefont
  {Arnold}}]{Gondrom1999}%
  \BibitemOpen
  \bibfield  {author} {\bibinfo {author} {\bibfnamefont {S.}~\bibnamefont
  {Gondrom}}, \bibinfo {author} {\bibfnamefont {J.}~\bibnamefont {Zhou}},
  \bibinfo {author} {\bibfnamefont {M.}~\bibnamefont {Maisl}}, \bibinfo
  {author} {\bibfnamefont {H.}~\bibnamefont {Reiter}}, \bibinfo {author}
  {\bibfnamefont {M.}~\bibnamefont {Kr{\"o}ning}}, \ and\ \bibinfo {author}
  {\bibfnamefont {W.}~\bibnamefont {Arnold}},\ }\href {\doibase
  10.1016/S0029-5493(98)00319-7} {\bibfield  {journal} {\bibinfo  {journal}
  {Nucl. Eng. Des.}\ }\textbf {\bibinfo {volume} {190}},\ \bibinfo {pages}
  {141} (\bibinfo {year} {1999})}\BibitemShut {NoStop}%
\bibitem [{\citenamefont {{Neculaes}}\ \emph {et~al.}(2014)\citenamefont
  {{Neculaes}}, \citenamefont {{Edic}}, \citenamefont {{Frontera}},
  \citenamefont {{Caiafa}}, \citenamefont {{Wang}},\ and\ \citenamefont {{De
  Man}}}]{Neculaes2014}%
  \BibitemOpen
  \bibfield  {author} {\bibinfo {author} {\bibfnamefont {V.~B.}\ \bibnamefont
  {{Neculaes}}}, \bibinfo {author} {\bibfnamefont {P.~M.}\ \bibnamefont
  {{Edic}}}, \bibinfo {author} {\bibfnamefont {M.}~\bibnamefont {{Frontera}}},
  \bibinfo {author} {\bibfnamefont {A.}~\bibnamefont {{Caiafa}}}, \bibinfo
  {author} {\bibfnamefont {G.}~\bibnamefont {{Wang}}}, \ and\ \bibinfo {author}
  {\bibfnamefont {B.}~\bibnamefont {{De Man}}},\ }\href {\doibase
  10.1109/ACCESS.2014.2363949} {\bibfield  {journal} {\bibinfo  {journal} {IEEE
  Access}\ }\textbf {\bibinfo {volume} {2}},\ \bibinfo {pages} {1568} (\bibinfo
  {year} {2014})}\BibitemShut {NoStop}%
\bibitem [{\citenamefont {Vigan\`{o}}\ \emph {et~al.}(2018)\citenamefont
  {Vigan\`{o}}, \citenamefont {Sarkissian}, \citenamefont {Herzog},
  \citenamefont {de~la Rochefoucauld}, \citenamefont {van Liere},\ and\
  \citenamefont {Batenburg}}]{Vigano2018}%
  \BibitemOpen
  \bibfield  {author} {\bibinfo {author} {\bibfnamefont {N.}~\bibnamefont
  {Vigan\`{o}}}, \bibinfo {author} {\bibfnamefont {H.~D.}\ \bibnamefont
  {Sarkissian}}, \bibinfo {author} {\bibfnamefont {C.}~\bibnamefont {Herzog}},
  \bibinfo {author} {\bibfnamefont {O.}~\bibnamefont {de~la Rochefoucauld}},
  \bibinfo {author} {\bibfnamefont {R.}~\bibnamefont {van Liere}}, \ and\
  \bibinfo {author} {\bibfnamefont {K.~J.}\ \bibnamefont {Batenburg}},\
  }\href@noop {} {\bibfield  {journal} {\bibinfo  {journal} {Opt. Express}\
  }\textbf {\bibinfo {volume} {26}},\ \bibinfo {pages} {22574} (\bibinfo {year}
  {2018})}\BibitemShut {NoStop}%
\bibitem [{\citenamefont {Vigan\`{o}}\ \emph {et~al.}(2019)\citenamefont
  {Vigan\`{o}}, \citenamefont {Gil}, \citenamefont {Herzog}, \citenamefont
  {de~la Rochefoucauld}, \citenamefont {van Liere},\ and\ \citenamefont
  {Batenburg}}]{Vigano2019}%
  \BibitemOpen
  \bibfield  {author} {\bibinfo {author} {\bibfnamefont {N.}~\bibnamefont
  {Vigan\`{o}}}, \bibinfo {author} {\bibfnamefont {P.~M.}\ \bibnamefont {Gil}},
  \bibinfo {author} {\bibfnamefont {C.}~\bibnamefont {Herzog}}, \bibinfo
  {author} {\bibfnamefont {O.}~\bibnamefont {de~la Rochefoucauld}}, \bibinfo
  {author} {\bibfnamefont {R.}~\bibnamefont {van Liere}}, \ and\ \bibinfo
  {author} {\bibfnamefont {K.~J.}\ \bibnamefont {Batenburg}},\ }\href@noop {}
  {\bibfield  {journal} {\bibinfo  {journal} {Opt. Express}\ }\textbf {\bibinfo
  {volume} {27}},\ \bibinfo {pages} {7834} (\bibinfo {year}
  {2019})}\BibitemShut {NoStop}%
\bibitem [{\citenamefont {Lippmann}(1908)}]{Lippmann1908}%
  \BibitemOpen
  \bibfield  {author} {\bibinfo {author} {\bibfnamefont {G.}~\bibnamefont
  {Lippmann}},\ }\href {\doibase 10.1051/jphystap:019080070082100} {\bibfield
  {journal} {\bibinfo  {journal} {J. Phys. Th{\'{e}}orique Appliqu{\'{e}}e}\
  }\textbf {\bibinfo {volume} {7}},\ \bibinfo {pages} {821} (\bibinfo {year}
  {1908})}\BibitemShut {NoStop}%
\bibitem [{\citenamefont {Adelson}\ and\ \citenamefont
  {Wang}(1992)}]{Adelson1992}%
  \BibitemOpen
  \bibfield  {author} {\bibinfo {author} {\bibfnamefont {E.}~\bibnamefont
  {Adelson}}\ and\ \bibinfo {author} {\bibfnamefont {J.}~\bibnamefont {Wang}},\
  }\href {\doibase 10.1109/34.121783} {\bibfield  {journal} {\bibinfo
  {journal} {IEEE Trans. Pattern Anal. Mach. Intell.}\ }\textbf {\bibinfo
  {volume} {14}},\ \bibinfo {pages} {99} (\bibinfo {year} {1992})}\BibitemShut
  {NoStop}%
\bibitem [{\citenamefont {Ng}\ \emph {et~al.}(2005)\citenamefont {Ng},
  \citenamefont {Levoy}, \citenamefont {Bredif}, \citenamefont {Duval},
  \citenamefont {Horowitz},\ and\ \citenamefont {Hanrahan}}]{Ng2005}%
  \BibitemOpen
  \bibfield  {author} {\bibinfo {author} {\bibfnamefont {R.}~\bibnamefont
  {Ng}}, \bibinfo {author} {\bibfnamefont {M.}~\bibnamefont {Levoy}}, \bibinfo
  {author} {\bibfnamefont {M.}~\bibnamefont {Bredif}}, \bibinfo {author}
  {\bibfnamefont {G.}~\bibnamefont {Duval}}, \bibinfo {author} {\bibfnamefont
  {M.}~\bibnamefont {Horowitz}}, \ and\ \bibinfo {author} {\bibfnamefont
  {P.}~\bibnamefont {Hanrahan}},\ }\href@noop {} {\enquote {\bibinfo {title}
  {{Light field photography with a hand-held plenoptic camera, CTSR
  2005-02}},}\ }\bibinfo {type} {Tech. Rep.}\ (\bibinfo  {institution}
  {Stanford},\ \bibinfo {year} {2005})\BibitemShut {NoStop}%
\bibitem [{\citenamefont {Georgiev}\ and\ \citenamefont
  {Lumsdaine}(2010)}]{Georgiev2010}%
  \BibitemOpen
  \bibfield  {author} {\bibinfo {author} {\bibfnamefont {T.}~\bibnamefont
  {Georgiev}}\ and\ \bibinfo {author} {\bibfnamefont {A.}~\bibnamefont
  {Lumsdaine}},\ }\href {\doibase 10.1117/1.3442712} {\bibfield  {journal}
  {\bibinfo  {journal} {J. Electron. Imaging}\ }\textbf {\bibinfo {volume}
  {19}},\ \bibinfo {pages} {021106} (\bibinfo {year} {2010})}\BibitemShut
  {NoStop}%
\bibitem [{\citenamefont {Lam}(2015)}]{Lam2015}%
  \BibitemOpen
  \bibfield  {author} {\bibinfo {author} {\bibfnamefont {E.~Y.}\ \bibnamefont
  {Lam}},\ }\href {\doibase 10.1364/JOSAA.32.002021} {\bibfield  {journal}
  {\bibinfo  {journal} {J. Opt. Soc. Am. A}\ }\textbf {\bibinfo {volume}
  {32}},\ \bibinfo {pages} {2021} (\bibinfo {year} {2015})}\BibitemShut
  {NoStop}%
\bibitem [{\citenamefont {Levoy}, \citenamefont {Zhang},\ and\ \citenamefont
  {Mcdowall}(2009)}]{Levoy2009}%
  \BibitemOpen
  \bibfield  {author} {\bibinfo {author} {\bibfnamefont {M.}~\bibnamefont
  {Levoy}}, \bibinfo {author} {\bibfnamefont {Z.}~\bibnamefont {Zhang}}, \ and\
  \bibinfo {author} {\bibfnamefont {I.}~\bibnamefont {Mcdowall}},\ }\href
  {\doibase 10.1111/j.1365-2818.2009.03195.x} {\bibfield  {journal} {\bibinfo
  {journal} {J. Microsc.}\ }\textbf {\bibinfo {volume} {235}},\ \bibinfo
  {pages} {144} (\bibinfo {year} {2009})}\BibitemShut {NoStop}%
\bibitem [{\citenamefont {D'Angelo}\ \emph {et~al.}(2016)\citenamefont
  {D'Angelo}, \citenamefont {Pepe}, \citenamefont {Garuccio},\ and\
  \citenamefont {Scarcelli}}]{Angelo2016}%
  \BibitemOpen
  \bibfield  {author} {\bibinfo {author} {\bibfnamefont {M.}~\bibnamefont
  {D'Angelo}}, \bibinfo {author} {\bibfnamefont {F.~V.}\ \bibnamefont {Pepe}},
  \bibinfo {author} {\bibfnamefont {A.}~\bibnamefont {Garuccio}}, \ and\
  \bibinfo {author} {\bibfnamefont {G.}~\bibnamefont {Scarcelli}},\ }\href
  {\doibase 10.1103/PhysRevLett.116.223602} {\bibfield  {journal} {\bibinfo
  {journal} {Phys. Rev. Lett.}\ }\textbf {\bibinfo {volume} {116}},\ \bibinfo
  {pages} {223602} (\bibinfo {year} {2016})}\BibitemShut {NoStop}%
\bibitem [{\citenamefont {Snigirev}\ \emph {et~al.}(1996)\citenamefont
  {Snigirev}, \citenamefont {Kohn}, \citenamefont {Snigireva},\ and\
  \citenamefont {Lengeler}}]{Snigirev1996}%
  \BibitemOpen
  \bibfield  {author} {\bibinfo {author} {\bibfnamefont {A.}~\bibnamefont
  {Snigirev}}, \bibinfo {author} {\bibfnamefont {V.}~\bibnamefont {Kohn}},
  \bibinfo {author} {\bibfnamefont {I.}~\bibnamefont {Snigireva}}, \ and\
  \bibinfo {author} {\bibfnamefont {B.}~\bibnamefont {Lengeler}},\ }\href@noop
  {} {\bibfield  {journal} {\bibinfo  {journal} {Nature}\ }\textbf {\bibinfo
  {volume} {384}},\ \bibinfo {pages} {49} (\bibinfo {year} {1996})}\BibitemShut
  {NoStop}%
\bibitem [{\citenamefont {Paganin}(2006)}]{paganin}%
  \BibitemOpen
  \bibfield  {author} {\bibinfo {author} {\bibfnamefont {D.}~\bibnamefont
  {Paganin}},\ }\href@noop {} {\emph {\bibinfo {title} {Coherent x-ray
  optics}}}\ (\bibinfo  {publisher} {Oxford University Press},\ \bibinfo {year}
  {2006})\BibitemShut {NoStop}%
\bibitem [{\citenamefont {Bilderback}, \citenamefont {Hoffman},\ and\
  \citenamefont {Thiel}(1994)}]{Bilderback1994}%
  \BibitemOpen
  \bibfield  {author} {\bibinfo {author} {\bibfnamefont {D.~H.}\ \bibnamefont
  {Bilderback}}, \bibinfo {author} {\bibfnamefont {S.~A.}\ \bibnamefont
  {Hoffman}}, \ and\ \bibinfo {author} {\bibfnamefont {D.~J.}\ \bibnamefont
  {Thiel}},\ }\href {\doibase 10.1126/science.8284671} {\bibfield  {journal}
  {\bibinfo  {journal} {Science}\ }\textbf {\bibinfo {volume} {263}},\ \bibinfo
  {pages} {201} (\bibinfo {year} {1994})}\BibitemShut {NoStop}%
\bibitem [{\citenamefont {Vincze}\ \emph {et~al.}(1998)\citenamefont {Vincze},
  \citenamefont {Janssens}, \citenamefont {Adams}, \citenamefont {Rindby},\
  and\ \citenamefont {Engstr{\"{o}}m}}]{Vincze1998}%
  \BibitemOpen
  \bibfield  {author} {\bibinfo {author} {\bibfnamefont {L.}~\bibnamefont
  {Vincze}}, \bibinfo {author} {\bibfnamefont {K.}~\bibnamefont {Janssens}},
  \bibinfo {author} {\bibfnamefont {F.}~\bibnamefont {Adams}}, \bibinfo
  {author} {\bibfnamefont {A.}~\bibnamefont {Rindby}}, \ and\ \bibinfo {author}
  {\bibfnamefont {P.}~\bibnamefont {Engstr{\"{o}}m}},\ }\href {\doibase
  10.1063/1.1149127} {\bibfield  {journal} {\bibinfo  {journal} {Rev. Sci.
  Instrum.}\ }\textbf {\bibinfo {volume} {69}},\ \bibinfo {pages} {3494}
  (\bibinfo {year} {1998})}\BibitemShut {NoStop}%
\bibitem [{\citenamefont {Chapman}, \citenamefont {Nugent},\ and\ \citenamefont
  {Wilkins}(1991)}]{Chapman1991}%
  \BibitemOpen
  \bibfield  {author} {\bibinfo {author} {\bibfnamefont {H.~N.}\ \bibnamefont
  {Chapman}}, \bibinfo {author} {\bibfnamefont {K.~A.}\ \bibnamefont {Nugent}},
  \ and\ \bibinfo {author} {\bibfnamefont {S.~W.}\ \bibnamefont {Wilkins}},\
  }\href {\doibase 10.1063/1.1142432} {\bibfield  {journal} {\bibinfo
  {journal} {Rev. Sci. Instrum.}\ }\textbf {\bibinfo {volume} {62}},\ \bibinfo
  {pages} {1542} (\bibinfo {year} {1991})}\BibitemShut {NoStop}%
\bibitem [{\citenamefont {Kumakhov}\ and\ \citenamefont
  {Komarov}(1990)}]{KumakhovPhysicsReports1990}%
  \BibitemOpen
  \bibfield  {author} {\bibinfo {author} {\bibfnamefont {M.}~\bibnamefont
  {Kumakhov}}\ and\ \bibinfo {author} {\bibfnamefont {F.}~\bibnamefont
  {Komarov}},\ }\href {\doibase 10.1016/0370-1573(90)90135-O} {\bibfield
  {journal} {\bibinfo  {journal} {Phys. Rep.}\ }\textbf {\bibinfo {volume}
  {191}},\ \bibinfo {pages} {289 } (\bibinfo {year} {1990})}\BibitemShut
  {NoStop}%
\bibitem [{\citenamefont {MacDonald}(2017)}]{MacDonald2017}%
  \BibitemOpen
  \bibfield  {author} {\bibinfo {author} {\bibfnamefont {C.~A.}\ \bibnamefont
  {MacDonald}},\ }\href@noop {} {\bibfield  {journal} {\bibinfo  {journal}
  {Annu. Rev. Mater. Res.}\ }\textbf {\bibinfo {volume} {47}},\ \bibinfo
  {pages} {115} (\bibinfo {year} {2017})}\BibitemShut {NoStop}%
\bibitem [{\citenamefont {Dabrowski}\ \emph {et~al.}(2013)\citenamefont
  {Dabrowski}, \citenamefont {Dul}, \citenamefont {Wrobel},\ and\ \citenamefont
  {Korecki}}]{Dabrowski2013APL}%
  \BibitemOpen
  \bibfield  {author} {\bibinfo {author} {\bibfnamefont {K.~M.}\ \bibnamefont
  {Dabrowski}}, \bibinfo {author} {\bibfnamefont {D.~T.}\ \bibnamefont {Dul}},
  \bibinfo {author} {\bibfnamefont {A.}~\bibnamefont {Wrobel}}, \ and\ \bibinfo
  {author} {\bibfnamefont {P.}~\bibnamefont {Korecki}},\ }\href
  {http://scitation.aip.org/content/aip/journal/apl/102/22/10.1063/1.4809583}
  {\bibfield  {journal} {\bibinfo  {journal} {Appl. Phys. Lett.}\ }\textbf
  {\bibinfo {volume} {102}},\ \bibinfo {eid} {224104} (\bibinfo {year}
  {2013})}\BibitemShut {NoStop}%
\bibitem [{\citenamefont {Fenimore}\ and\ \citenamefont
  {Cannon}(1978)}]{Fenimore1978}%
  \BibitemOpen
  \bibfield  {author} {\bibinfo {author} {\bibfnamefont {E.~E.}\ \bibnamefont
  {Fenimore}}\ and\ \bibinfo {author} {\bibfnamefont {T.~M.}\ \bibnamefont
  {Cannon}},\ }\href {\doibase 10.1364/AO.17.000337} {\bibfield  {journal}
  {\bibinfo  {journal} {Appl. Opt.}\ }\textbf {\bibinfo {volume} {17}},\
  \bibinfo {pages} {337} (\bibinfo {year} {1978})}\BibitemShut {NoStop}%
\bibitem [{\citenamefont {Nadjmi}\ \emph {et~al.}(1980)\citenamefont {Nadjmi},
  \citenamefont {Weiss}, \citenamefont {Klotz},\ and\ \citenamefont
  {Linde}}]{Nadjmi1980}%
  \BibitemOpen
  \bibfield  {author} {\bibinfo {author} {\bibfnamefont {M.}~\bibnamefont
  {Nadjmi}}, \bibinfo {author} {\bibfnamefont {H.}~\bibnamefont {Weiss}},
  \bibinfo {author} {\bibfnamefont {E.}~\bibnamefont {Klotz}}, \ and\ \bibinfo
  {author} {\bibfnamefont {R.}~\bibnamefont {Linde}},\ }\href {\doibase
  10.1007/BF00342384} {\bibfield  {journal} {\bibinfo  {journal}
  {Neuroradiology}\ }\textbf {\bibinfo {volume} {19}},\ \bibinfo {pages} {113}
  (\bibinfo {year} {1980})}\BibitemShut {NoStop}%
\bibitem [{\citenamefont {Dabrowski}, \citenamefont {Dul},\ and\ \citenamefont
  {Korecki}(2013)}]{DabrowskiOE2013}%
  \BibitemOpen
  \bibfield  {author} {\bibinfo {author} {\bibfnamefont {K.~M.}\ \bibnamefont
  {Dabrowski}}, \bibinfo {author} {\bibfnamefont {D.~T.}\ \bibnamefont {Dul}},
  \ and\ \bibinfo {author} {\bibfnamefont {P.}~\bibnamefont {Korecki}},\ }\href
  {\doibase 10.1364/OE.21.002920} {\bibfield  {journal} {\bibinfo  {journal}
  {Opt. Express}\ }\textbf {\bibinfo {volume} {21}},\ \bibinfo {pages} {2920}
  (\bibinfo {year} {2013})}\BibitemShut {NoStop}%
\bibitem [{\citenamefont {Sowa}, \citenamefont {Last},\ and\ \citenamefont
  {Korecki}(2017)}]{Sowa2017}%
  \BibitemOpen
  \bibfield  {author} {\bibinfo {author} {\bibfnamefont {K.~M.}\ \bibnamefont
  {Sowa}}, \bibinfo {author} {\bibfnamefont {A.}~\bibnamefont {Last}}, \ and\
  \bibinfo {author} {\bibfnamefont {P.}~\bibnamefont {Korecki}},\ }\href@noop
  {} {\bibfield  {journal} {\bibinfo  {journal} {Sci. Rep.}\ }\textbf {\bibinfo
  {volume} {7}},\ \bibinfo {pages} {44944} (\bibinfo {year}
  {2017})}\BibitemShut {NoStop}%
\bibitem [{\citenamefont {Sowa}, \citenamefont {Jany},\ and\ \citenamefont
  {Korecki}(2018)}]{Sowa2018}%
  \BibitemOpen
  \bibfield  {author} {\bibinfo {author} {\bibfnamefont {K.~M.}\ \bibnamefont
  {Sowa}}, \bibinfo {author} {\bibfnamefont {B.~R.}\ \bibnamefont {Jany}}, \
  and\ \bibinfo {author} {\bibfnamefont {P.}~\bibnamefont {Korecki}},\
  }\href@noop {} {\bibfield  {journal} {\bibinfo  {journal} {Optica}\ }\textbf
  {\bibinfo {volume} {5}},\ \bibinfo {pages} {577} (\bibinfo {year}
  {2018})}\BibitemShut {NoStop}%
\bibitem [{\citenamefont {Korecki}\ \emph {et~al.}(2016)\citenamefont
  {Korecki}, \citenamefont {Sowa}, \citenamefont {Jany},\ and\ \citenamefont
  {Krok}}]{Korecki2017}%
  \BibitemOpen
  \bibfield  {author} {\bibinfo {author} {\bibfnamefont {P.}~\bibnamefont
  {Korecki}}, \bibinfo {author} {\bibfnamefont {K.~M.}\ \bibnamefont {Sowa}},
  \bibinfo {author} {\bibfnamefont {B.~R.}\ \bibnamefont {Jany}}, \ and\
  \bibinfo {author} {\bibfnamefont {F.}~\bibnamefont {Krok}},\ }\href {\doibase
  10.1103/PhysRevLett.116.233902} {\bibfield  {journal} {\bibinfo  {journal}
  {Phys. Rev. Lett.}\ }\textbf {\bibinfo {volume} {116}},\ \bibinfo {pages}
  {233902} (\bibinfo {year} {2016})}\BibitemShut {NoStop}%
\bibitem [{\citenamefont {Korecki}, \citenamefont {Roszczynialski},\ and\
  \citenamefont {Sowa}(2015)}]{Korecki2015}%
  \BibitemOpen
  \bibfield  {author} {\bibinfo {author} {\bibfnamefont {P.}~\bibnamefont
  {Korecki}}, \bibinfo {author} {\bibfnamefont {T.~P.}\ \bibnamefont
  {Roszczynialski}}, \ and\ \bibinfo {author} {\bibfnamefont {K.~M.}\
  \bibnamefont {Sowa}},\ }\href@noop {} {\bibfield  {journal} {\bibinfo
  {journal} {Opt. Express}\ }\textbf {\bibinfo {volume} {23}},\ \bibinfo
  {pages} {8749} (\bibinfo {year} {2015})}\BibitemShut {NoStop}%
\bibitem [{\citenamefont {Snigirev}\ \emph {et~al.}(1995)\citenamefont
  {Snigirev}, \citenamefont {Snigireva}, \citenamefont {Kohn}, \citenamefont
  {Kuznetsov},\ and\ \citenamefont {Schelokov}}]{Snigirev1995}%
  \BibitemOpen
  \bibfield  {author} {\bibinfo {author} {\bibfnamefont {A.}~\bibnamefont
  {Snigirev}}, \bibinfo {author} {\bibfnamefont {I.}~\bibnamefont {Snigireva}},
  \bibinfo {author} {\bibfnamefont {V.}~\bibnamefont {Kohn}}, \bibinfo {author}
  {\bibfnamefont {S.}~\bibnamefont {Kuznetsov}}, \ and\ \bibinfo {author}
  {\bibfnamefont {I.}~\bibnamefont {Schelokov}},\ }\href {\doibase
  10.1063/1.1146073} {\bibfield  {journal} {\bibinfo  {journal} {Rev. Sci.
  Instrum.}\ }\textbf {\bibinfo {volume} {66}},\ \bibinfo {pages} {5486}
  (\bibinfo {year} {1995})}\BibitemShut {NoStop}%
\bibitem [{\citenamefont {Cloetens}\ \emph {et~al.}(1996)\citenamefont
  {Cloetens}, \citenamefont {Barrett}, \citenamefont {Baruchel}, \citenamefont
  {Guigay},\ and\ \citenamefont {Schlenker}}]{Cloetens1996}%
  \BibitemOpen
  \bibfield  {author} {\bibinfo {author} {\bibfnamefont {P.}~\bibnamefont
  {Cloetens}}, \bibinfo {author} {\bibfnamefont {R.}~\bibnamefont {Barrett}},
  \bibinfo {author} {\bibfnamefont {J.}~\bibnamefont {Baruchel}}, \bibinfo
  {author} {\bibfnamefont {J.-P.}\ \bibnamefont {Guigay}}, \ and\ \bibinfo
  {author} {\bibfnamefont {M.}~\bibnamefont {Schlenker}},\ }\href
  {http://stacks.iop.org/0022-3727/29/i=1/a=023} {\bibfield  {journal}
  {\bibinfo  {journal} {J. Phys. D: Appl. Phys.}\ }\textbf {\bibinfo {volume}
  {29}},\ \bibinfo {pages} {133} (\bibinfo {year} {1996})}\BibitemShut
  {NoStop}%
\bibitem [{\citenamefont {Pogany}, \citenamefont {Gao},\ and\ \citenamefont
  {Wilkins}(1997)}]{Pogany1997}%
  \BibitemOpen
  \bibfield  {author} {\bibinfo {author} {\bibfnamefont {A.}~\bibnamefont
  {Pogany}}, \bibinfo {author} {\bibfnamefont {D.}~\bibnamefont {Gao}}, \ and\
  \bibinfo {author} {\bibfnamefont {S.~W.}\ \bibnamefont {Wilkins}},\
  }\href@noop {} {\bibfield  {journal} {\bibinfo  {journal} {Review of
  Scientific Instruments}\ }\textbf {\bibinfo {volume} {68}},\ \bibinfo {pages}
  {2774} (\bibinfo {year} {1997})}\BibitemShut {NoStop}%
\bibitem [{\citenamefont {Gureyev}\ \emph {et~al.}(2001)\citenamefont
  {Gureyev}, \citenamefont {Mayo}, \citenamefont {Wilkins}, \citenamefont
  {Paganin},\ and\ \citenamefont {Stevenson}}]{Gureyev2001}%
  \BibitemOpen
  \bibfield  {author} {\bibinfo {author} {\bibfnamefont {T.~E.}\ \bibnamefont
  {Gureyev}}, \bibinfo {author} {\bibfnamefont {S.}~\bibnamefont {Mayo}},
  \bibinfo {author} {\bibfnamefont {S.~W.}\ \bibnamefont {Wilkins}}, \bibinfo
  {author} {\bibfnamefont {D.}~\bibnamefont {Paganin}}, \ and\ \bibinfo
  {author} {\bibfnamefont {A.~W.}\ \bibnamefont {Stevenson}},\ }\href {\doibase
  10.1103/PhysRevLett.86.5827} {\bibfield  {journal} {\bibinfo  {journal}
  {Phys. Rev. Lett.}\ }\textbf {\bibinfo {volume} {86}},\ \bibinfo {pages}
  {5827} (\bibinfo {year} {2001})}\BibitemShut {NoStop}%
\bibitem [{\citenamefont {Grant}(1972)}]{Grant1972}%
  \BibitemOpen
  \bibfield  {author} {\bibinfo {author} {\bibfnamefont {D.~G.}\ \bibnamefont
  {Grant}},\ }\href {\doibase 10.1109/TBME.1972.324154} {\bibfield  {journal}
  {\bibinfo  {journal} {IEEE Trans. Biomed. Eng.}\ }\textbf {\bibinfo {volume}
  {19}},\ \bibinfo {pages} {20} (\bibinfo {year} {1972})}\BibitemShut {NoStop}%
\bibitem [{\citenamefont {Gureyev}\ \emph {et~al.}(2009)\citenamefont
  {Gureyev}, \citenamefont {Mayo}, \citenamefont {Myers}, \citenamefont
  {Nesterets}, \citenamefont {Paganin}, \citenamefont {Pogany}, \citenamefont
  {Stevenson},\ and\ \citenamefont {Wilkins}}]{Gureyev2009}%
  \BibitemOpen
  \bibfield  {author} {\bibinfo {author} {\bibfnamefont {T.~E.}\ \bibnamefont
  {Gureyev}}, \bibinfo {author} {\bibfnamefont {S.~C.}\ \bibnamefont {Mayo}},
  \bibinfo {author} {\bibfnamefont {D.~E.}\ \bibnamefont {Myers}}, \bibinfo
  {author} {\bibfnamefont {Y.}~\bibnamefont {Nesterets}}, \bibinfo {author}
  {\bibfnamefont {D.~M.}\ \bibnamefont {Paganin}}, \bibinfo {author}
  {\bibfnamefont {A.}~\bibnamefont {Pogany}}, \bibinfo {author} {\bibfnamefont
  {A.~W.}\ \bibnamefont {Stevenson}}, \ and\ \bibinfo {author} {\bibfnamefont
  {S.~W.}\ \bibnamefont {Wilkins}},\ }\href {\doibase 10.1063/1.3115402}
  {\bibfield  {journal} {\bibinfo  {journal} {Journal of Applied Physics}\
  }\textbf {\bibinfo {volume} {105}},\ \bibinfo {pages} {102005} (\bibinfo
  {year} {2009})}\BibitemShut {NoStop}%
\bibitem [{\citenamefont {Franceschi}\ and\ \citenamefont
  {Nakata}(2005)}]{Franceschi2005}%
  \BibitemOpen
  \bibfield  {author} {\bibinfo {author} {\bibfnamefont {V.~R.}\ \bibnamefont
  {Franceschi}}\ and\ \bibinfo {author} {\bibfnamefont {P.~A.}\ \bibnamefont
  {Nakata}},\ }\href {\doibase 10.1146/annurev.arplant.56.032604.144106}
  {\bibfield  {journal} {\bibinfo  {journal} {Annu. Rev. Plant Biol.}\ }\textbf
  {\bibinfo {volume} {56}},\ \bibinfo {pages} {41} (\bibinfo {year}
  {2005})}\BibitemShut {NoStop}%
\bibitem [{\citenamefont {da~Costa}\ \emph {et~al.}(2009)\citenamefont
  {da~Costa}, \citenamefont {Tronto}, \citenamefont {Constantino},
  \citenamefont {Fonseca}, \citenamefont {Oliveira},\ and\ \citenamefont
  {da~Costa}}]{Costa2009}%
  \BibitemOpen
  \bibfield  {author} {\bibinfo {author} {\bibfnamefont {L.~M.}\ \bibnamefont
  {da~Costa}}, \bibinfo {author} {\bibfnamefont {J.}~\bibnamefont {Tronto}},
  \bibinfo {author} {\bibfnamefont {V.~R.~L.}\ \bibnamefont {Constantino}},
  \bibinfo {author} {\bibfnamefont {M.~K.~A.}\ \bibnamefont {Fonseca}},
  \bibinfo {author} {\bibfnamefont {A.~P.}\ \bibnamefont {Oliveira}}, \ and\
  \bibinfo {author} {\bibfnamefont {M.~R.}\ \bibnamefont {da~Costa}},\ }\href
  {\doibase 10.1590/S0100-06832009000300025} {\bibfield  {journal} {\bibinfo
  {journal} {Rev. Bras. Ci{\^{e}}ncia do Solo}\ }\textbf {\bibinfo {volume}
  {33}},\ \bibinfo {pages} {729} (\bibinfo {year} {2009})}\BibitemShut
  {NoStop}%
\bibitem [{\citenamefont {Ackerfield}(2001)}]{Ackerfield2001}%
  \BibitemOpen
  \bibfield  {author} {\bibinfo {author} {\bibfnamefont {J.}~\bibnamefont
  {Ackerfield}},\ }\href {\doibase 10.1017/S0960428601000622} {\bibfield
  {journal} {\bibinfo  {journal} {Edinburgh J. Bot.}\ }\textbf {\bibinfo
  {volume} {58}},\ \bibinfo {pages} {259} (\bibinfo {year} {2001})}\BibitemShut
  {NoStop}%
\bibitem [{\citenamefont {Kirz}, \citenamefont {Jacobsen},\ and\ \citenamefont
  {Howells}(1995)}]{kirz1995}%
  \BibitemOpen
  \bibfield  {author} {\bibinfo {author} {\bibfnamefont {J.}~\bibnamefont
  {Kirz}}, \bibinfo {author} {\bibfnamefont {C.}~\bibnamefont {Jacobsen}}, \
  and\ \bibinfo {author} {\bibfnamefont {M.}~\bibnamefont {Howells}},\ }\href
  {\doibase 10.1017/S0033583500003139} {\bibfield  {journal} {\bibinfo
  {journal} {Quarterly Reviews of Biophysics}\ }\textbf {\bibinfo {volume}
  {28}},\ \bibinfo {pages} {33–130} (\bibinfo {year} {1995})}\BibitemShut
  {NoStop}%
\bibitem [{\citenamefont {Daniels}\ and\ \citenamefont
  {Brennan}(1996)}]{DANIELS1996}%
  \BibitemOpen
  \bibfield  {author} {\bibinfo {author} {\bibfnamefont {S.}~\bibnamefont
  {Daniels}}\ and\ \bibinfo {author} {\bibfnamefont {P.}~\bibnamefont
  {Brennan}},\ }\href {\doibase https://doi.org/10.1016/S1078-8174(96)90002-4}
  {\bibfield  {journal} {\bibinfo  {journal} {Radiography}\ }\textbf {\bibinfo
  {volume} {2}},\ \bibinfo {pages} {99 } (\bibinfo {year} {1996})}\BibitemShut
  {NoStop}%
\bibitem [{\citenamefont {Vedantham}\ \emph {et~al.}(2015)\citenamefont
  {Vedantham}, \citenamefont {Karellas}, \citenamefont {Vijayaraghavan},\ and\
  \citenamefont {Kopans}}]{Vedantham2015}%
  \BibitemOpen
  \bibfield  {author} {\bibinfo {author} {\bibfnamefont {S.}~\bibnamefont
  {Vedantham}}, \bibinfo {author} {\bibfnamefont {A.}~\bibnamefont {Karellas}},
  \bibinfo {author} {\bibfnamefont {G.~R.}\ \bibnamefont {Vijayaraghavan}}, \
  and\ \bibinfo {author} {\bibfnamefont {D.~B.}\ \bibnamefont {Kopans}},\
  }\href {\doibase 10.1148/radiol.2015141303} {\bibfield  {journal} {\bibinfo
  {journal} {Radiology}\ }\textbf {\bibinfo {volume} {277}},\ \bibinfo {pages}
  {663} (\bibinfo {year} {2015})},\ \bibinfo {note} {pMID:
  26599926}\BibitemShut {NoStop}%
\bibitem [{\citenamefont {Paganin}\ \emph {et~al.}(2002)\citenamefont
  {Paganin}, \citenamefont {Mayo}, \citenamefont {Gureyev}, \citenamefont
  {Miller},\ and\ \citenamefont {Wilkins}}]{Paganin2002}%
  \BibitemOpen
  \bibfield  {author} {\bibinfo {author} {\bibfnamefont {D.}~\bibnamefont
  {Paganin}}, \bibinfo {author} {\bibfnamefont {S.~C.}\ \bibnamefont {Mayo}},
  \bibinfo {author} {\bibfnamefont {T.~E.}\ \bibnamefont {Gureyev}}, \bibinfo
  {author} {\bibfnamefont {P.~R.}\ \bibnamefont {Miller}}, \ and\ \bibinfo
  {author} {\bibfnamefont {S.~W.}\ \bibnamefont {Wilkins}},\ }\href {\doibase
  10.1046/j.1365-2818.2002.01010.x} {\bibfield  {journal} {\bibinfo  {journal}
  {J. Microsc.}\ }\textbf {\bibinfo {volume} {206}},\ \bibinfo {pages} {33}
  (\bibinfo {year} {2002})}\BibitemShut {NoStop}%
\bibitem [{\citenamefont {Thompson}\ \emph {et~al.}(2019)\citenamefont
  {Thompson}, \citenamefont {Nesterets}, \citenamefont {Pavlov},\ and\
  \citenamefont {Gureyev}}]{Thompson2019}%
  \BibitemOpen
  \bibfield  {author} {\bibinfo {author} {\bibfnamefont {D.~A.}\ \bibnamefont
  {Thompson}}, \bibinfo {author} {\bibfnamefont {Y.~I.}\ \bibnamefont
  {Nesterets}}, \bibinfo {author} {\bibfnamefont {K.~M.}\ \bibnamefont
  {Pavlov}}, \ and\ \bibinfo {author} {\bibfnamefont {T.~E.}\ \bibnamefont
  {Gureyev}},\ }\href {\doibase 10.1107/S1600577519002133} {\bibfield
  {journal} {\bibinfo  {journal} {Journal of Synchrotron Radiation}\ }\textbf
  {\bibinfo {volume} {26}},\ \bibinfo {pages} {825} (\bibinfo {year}
  {2019})}\BibitemShut {NoStop}%
\bibitem [{\citenamefont {{Du}}\ \emph {et~al.}(2019)\citenamefont {{Du}},
  \citenamefont {{Nashed}}, \citenamefont {{Kandel}}, \citenamefont
  {{Gursoy}},\ and\ \citenamefont {{Jacobsen}}}]{Ming2019}%
  \BibitemOpen
  \bibfield  {author} {\bibinfo {author} {\bibfnamefont {M.}~\bibnamefont
  {{Du}}}, \bibinfo {author} {\bibfnamefont {Y.~S.~G.}\ \bibnamefont
  {{Nashed}}}, \bibinfo {author} {\bibfnamefont {S.}~\bibnamefont {{Kandel}}},
  \bibinfo {author} {\bibfnamefont {D.}~\bibnamefont {{Gursoy}}}, \ and\
  \bibinfo {author} {\bibfnamefont {C.}~\bibnamefont {{Jacobsen}}},\
  }\href@noop {} {\bibfield  {journal} {\bibinfo  {journal} {arXiv e-prints}\
  ,\ \bibinfo {eid} {arXiv:1905.10433}} (\bibinfo {year} {2019})},\ \Eprint
  {http://arxiv.org/abs/1905.10433} {arXiv:1905.10433 [eess.IV]} \BibitemShut
  {NoStop}%
\bibitem [{\citenamefont {Dehlinger}\ \emph {et~al.}(2013)\citenamefont
  {Dehlinger}, \citenamefont {Fauquet}, \citenamefont {Jandard}, \citenamefont
  {Bjeoumikhov}, \citenamefont {Bjeoumikhova}, \citenamefont {Gubzhokov},
  \citenamefont {Erko}, \citenamefont {Zizak}, \citenamefont {Pailharey},
  \citenamefont {Ferrero}, \citenamefont {Dahmani},\ and\ \citenamefont
  {Tonneau}}]{Dehlinger2013}%
  \BibitemOpen
  \bibfield  {author} {\bibinfo {author} {\bibfnamefont {M.}~\bibnamefont
  {Dehlinger}}, \bibinfo {author} {\bibfnamefont {C.}~\bibnamefont {Fauquet}},
  \bibinfo {author} {\bibfnamefont {F.}~\bibnamefont {Jandard}}, \bibinfo
  {author} {\bibfnamefont {A.}~\bibnamefont {Bjeoumikhov}}, \bibinfo {author}
  {\bibfnamefont {S.}~\bibnamefont {Bjeoumikhova}}, \bibinfo {author}
  {\bibfnamefont {R.}~\bibnamefont {Gubzhokov}}, \bibinfo {author}
  {\bibfnamefont {A.}~\bibnamefont {Erko}}, \bibinfo {author} {\bibfnamefont
  {I.}~\bibnamefont {Zizak}}, \bibinfo {author} {\bibfnamefont
  {D.}~\bibnamefont {Pailharey}}, \bibinfo {author} {\bibfnamefont
  {S.}~\bibnamefont {Ferrero}}, \bibinfo {author} {\bibfnamefont
  {B.}~\bibnamefont {Dahmani}}, \ and\ \bibinfo {author} {\bibfnamefont
  {D.}~\bibnamefont {Tonneau}},\ }\href@noop {} {\bibfield  {journal} {\bibinfo
   {journal} {X-Ray Spectrometry}\ }\textbf {\bibinfo {volume} {42}},\ \bibinfo
  {pages} {456} (\bibinfo {year} {2013})}\BibitemShut {NoStop}%
\bibitem [{\citenamefont {Brombal}\ \emph {et~al.}(2019)\citenamefont
  {Brombal}, \citenamefont {Kallon}, \citenamefont {Jiang}, \citenamefont
  {Savvidis}, \citenamefont {De~Coppi}, \citenamefont {Urbani}, \citenamefont
  {Forty}, \citenamefont {Chambers}, \citenamefont {Longo}, \citenamefont
  {Olivo},\ and\ \citenamefont {Endrizzi}}]{Brombal2019}%
  \BibitemOpen
  \bibfield  {author} {\bibinfo {author} {\bibfnamefont {L.}~\bibnamefont
  {Brombal}}, \bibinfo {author} {\bibfnamefont {G.}~\bibnamefont {Kallon}},
  \bibinfo {author} {\bibfnamefont {J.}~\bibnamefont {Jiang}}, \bibinfo
  {author} {\bibfnamefont {S.}~\bibnamefont {Savvidis}}, \bibinfo {author}
  {\bibfnamefont {P.}~\bibnamefont {De~Coppi}}, \bibinfo {author}
  {\bibfnamefont {L.}~\bibnamefont {Urbani}}, \bibinfo {author} {\bibfnamefont
  {E.}~\bibnamefont {Forty}}, \bibinfo {author} {\bibfnamefont
  {R.}~\bibnamefont {Chambers}}, \bibinfo {author} {\bibfnamefont
  {R.}~\bibnamefont {Longo}}, \bibinfo {author} {\bibfnamefont
  {A.}~\bibnamefont {Olivo}}, \ and\ \bibinfo {author} {\bibfnamefont
  {M.}~\bibnamefont {Endrizzi}},\ }\href@noop {} {\bibfield  {journal}
  {\bibinfo  {journal} {Phys. Rev. Applied}\ }\textbf {\bibinfo {volume}
  {11}},\ \bibinfo {pages} {034004} (\bibinfo {year} {2019})}\BibitemShut
  {NoStop}%
\bibitem [{\citenamefont {Bauer}\ \emph {et~al.}(2018)\citenamefont {Bauer},
  \citenamefont {Lindqvist}, \citenamefont {Förste}, \citenamefont
  {Lundström}, \citenamefont {Hansson}, \citenamefont {Thiel}, \citenamefont
  {Bjeoumikhova}, \citenamefont {Grötzsch}, \citenamefont {Malzer},
  \citenamefont {Kanngießer},\ and\ \citenamefont {Mantouvalou}}]{Bauer2018}%
  \BibitemOpen
  \bibfield  {author} {\bibinfo {author} {\bibfnamefont {L.}~\bibnamefont
  {Bauer}}, \bibinfo {author} {\bibfnamefont {M.}~\bibnamefont {Lindqvist}},
  \bibinfo {author} {\bibfnamefont {F.}~\bibnamefont {Förste}}, \bibinfo
  {author} {\bibfnamefont {U.}~\bibnamefont {Lundström}}, \bibinfo {author}
  {\bibfnamefont {B.}~\bibnamefont {Hansson}}, \bibinfo {author} {\bibfnamefont
  {M.}~\bibnamefont {Thiel}}, \bibinfo {author} {\bibfnamefont
  {S.}~\bibnamefont {Bjeoumikhova}}, \bibinfo {author} {\bibfnamefont
  {D.}~\bibnamefont {Grötzsch}}, \bibinfo {author} {\bibfnamefont
  {W.}~\bibnamefont {Malzer}}, \bibinfo {author} {\bibfnamefont
  {B.}~\bibnamefont {Kanngießer}}, \ and\ \bibinfo {author} {\bibfnamefont
  {I.}~\bibnamefont {Mantouvalou}},\ }\href {\doibase 10.1039/C8JA00174J}
  {\bibfield  {journal} {\bibinfo  {journal} {J. Anal. At. Spectrom.}\ }\textbf
  {\bibinfo {volume} {33}},\ \bibinfo {pages} {1552} (\bibinfo {year}
  {2018})}\BibitemShut {NoStop}%
\bibitem [{\citenamefont {Graf}\ \emph {et~al.}(2018)\citenamefont {Graf},
  \citenamefont {Stuerzer}, \citenamefont {Ott}, \citenamefont {Mrosek},
  \citenamefont {Benning}, \citenamefont {Noll}, \citenamefont {Durst},
  \citenamefont {Radcliffe},\ and\ \citenamefont {Michaelsen}}]{Graf2018}%
  \BibitemOpen
  \bibfield  {author} {\bibinfo {author} {\bibfnamefont {J.}~\bibnamefont
  {Graf}}, \bibinfo {author} {\bibfnamefont {T.}~\bibnamefont {Stuerzer}},
  \bibinfo {author} {\bibfnamefont {H.}~\bibnamefont {Ott}}, \bibinfo {author}
  {\bibfnamefont {M.}~\bibnamefont {Mrosek}}, \bibinfo {author} {\bibfnamefont
  {M.}~\bibnamefont {Benning}}, \bibinfo {author} {\bibfnamefont
  {B.}~\bibnamefont {Noll}}, \bibinfo {author} {\bibfnamefont {R.}~\bibnamefont
  {Durst}}, \bibinfo {author} {\bibfnamefont {P.}~\bibnamefont {Radcliffe}}, \
  and\ \bibinfo {author} {\bibfnamefont {C.}~\bibnamefont {Michaelsen}},\
  }\href {\doibase 10.1107/S0108767318097994} {\bibfield  {journal} {\bibinfo
  {journal} {Acta Crystallographica Section A}\ }\textbf {\bibinfo {volume}
  {74}},\ \bibinfo {pages} {a200} (\bibinfo {year} {2018})}\BibitemShut
  {NoStop}%
\bibitem [{\citenamefont {Casanas}\ \emph {et~al.}(2016)\citenamefont
  {Casanas}, \citenamefont {Warshamanage}, \citenamefont {Finke}, \citenamefont
  {Panepucci}, \citenamefont {Olieric}, \citenamefont {N{\"{o}}ll},
  \citenamefont {Tamp{\'{e}}}, \citenamefont {Brandstetter}, \citenamefont
  {F{\"{o}}rster}, \citenamefont {Mueller}, \citenamefont {Schulze-Briese},
  \citenamefont {Bunk},\ and\ \citenamefont {Wang}}]{Casanas2016}%
  \BibitemOpen
  \bibfield  {author} {\bibinfo {author} {\bibfnamefont {A.}~\bibnamefont
  {Casanas}}, \bibinfo {author} {\bibfnamefont {R.}~\bibnamefont
  {Warshamanage}}, \bibinfo {author} {\bibfnamefont {A.~D.}\ \bibnamefont
  {Finke}}, \bibinfo {author} {\bibfnamefont {E.}~\bibnamefont {Panepucci}},
  \bibinfo {author} {\bibfnamefont {V.}~\bibnamefont {Olieric}}, \bibinfo
  {author} {\bibfnamefont {A.}~\bibnamefont {N{\"{o}}ll}}, \bibinfo {author}
  {\bibfnamefont {R.}~\bibnamefont {Tamp{\'{e}}}}, \bibinfo {author}
  {\bibfnamefont {S.}~\bibnamefont {Brandstetter}}, \bibinfo {author}
  {\bibfnamefont {A.}~\bibnamefont {F{\"{o}}rster}}, \bibinfo {author}
  {\bibfnamefont {M.}~\bibnamefont {Mueller}}, \bibinfo {author} {\bibfnamefont
  {C.}~\bibnamefont {Schulze-Briese}}, \bibinfo {author} {\bibfnamefont
  {O.}~\bibnamefont {Bunk}}, \ and\ \bibinfo {author} {\bibfnamefont
  {M.}~\bibnamefont {Wang}},\ }\href {\doibase 10.1107/S2059798316012304}
  {\bibfield  {journal} {\bibinfo  {journal} {Acta Crystallographica Section
  D}\ }\textbf {\bibinfo {volume} {72}},\ \bibinfo {pages} {1036} (\bibinfo
  {year} {2016})}\BibitemShut {NoStop}%
\bibitem [{\citenamefont {Villanueva-Perez}\ \emph {et~al.}(2018)\citenamefont
  {Villanueva-Perez}, \citenamefont {Pedrini}, \citenamefont {Mokso},
  \citenamefont {Vagovic}, \citenamefont {Guzenko}, \citenamefont {Leake},
  \citenamefont {Willmott}, \citenamefont {Oberta}, \citenamefont {David},
  \citenamefont {Chapman},\ and\ \citenamefont
  {Stampanoni}}]{Villanueva-Perez2018}%
  \BibitemOpen
  \bibfield  {author} {\bibinfo {author} {\bibfnamefont {P.}~\bibnamefont
  {Villanueva-Perez}}, \bibinfo {author} {\bibfnamefont {B.}~\bibnamefont
  {Pedrini}}, \bibinfo {author} {\bibfnamefont {R.}~\bibnamefont {Mokso}},
  \bibinfo {author} {\bibfnamefont {P.}~\bibnamefont {Vagovic}}, \bibinfo
  {author} {\bibfnamefont {V.~A.}\ \bibnamefont {Guzenko}}, \bibinfo {author}
  {\bibfnamefont {S.~J.}\ \bibnamefont {Leake}}, \bibinfo {author}
  {\bibfnamefont {P.~R.}\ \bibnamefont {Willmott}}, \bibinfo {author}
  {\bibfnamefont {P.}~\bibnamefont {Oberta}}, \bibinfo {author} {\bibfnamefont
  {C.}~\bibnamefont {David}}, \bibinfo {author} {\bibfnamefont {H.~N.}\
  \bibnamefont {Chapman}}, \ and\ \bibinfo {author} {\bibfnamefont
  {M.}~\bibnamefont {Stampanoni}},\ }\href {\doibase 10.1364/OPTICA.5.001521}
  {\bibfield  {journal} {\bibinfo  {journal} {Optica}\ }\textbf {\bibinfo
  {volume} {5}},\ \bibinfo {pages} {1521} (\bibinfo {year} {2018})}\BibitemShut
  {NoStop}%
\bibitem [{\citenamefont {Duarte}\ \emph {et~al.}(2019)\citenamefont {Duarte},
  \citenamefont {Cassin}, \citenamefont {Huijts}, \citenamefont {Iwan},
  \citenamefont {Fortuna}, \citenamefont {Delbecq}, \citenamefont {Chapman},
  \citenamefont {Fajardo}, \citenamefont {Kovacev}, \citenamefont {Boutu},\
  and\ \citenamefont {Merdji}}]{Duarte2019}%
  \BibitemOpen
  \bibfield  {author} {\bibinfo {author} {\bibfnamefont {J.}~\bibnamefont
  {Duarte}}, \bibinfo {author} {\bibfnamefont {R.}~\bibnamefont {Cassin}},
  \bibinfo {author} {\bibfnamefont {J.}~\bibnamefont {Huijts}}, \bibinfo
  {author} {\bibfnamefont {B.}~\bibnamefont {Iwan}}, \bibinfo {author}
  {\bibfnamefont {F.}~\bibnamefont {Fortuna}}, \bibinfo {author} {\bibfnamefont
  {L.}~\bibnamefont {Delbecq}}, \bibinfo {author} {\bibfnamefont
  {H.}~\bibnamefont {Chapman}}, \bibinfo {author} {\bibfnamefont
  {M.}~\bibnamefont {Fajardo}}, \bibinfo {author} {\bibfnamefont
  {M.}~\bibnamefont {Kovacev}}, \bibinfo {author} {\bibfnamefont
  {W.}~\bibnamefont {Boutu}}, \ and\ \bibinfo {author} {\bibfnamefont
  {H.}~\bibnamefont {Merdji}},\ }\href {\doibase 10.1038/s41566-019-0419-1}
  {\bibfield  {journal} {\bibinfo  {journal} {Nature Photonics}\ }\textbf
  {\bibinfo {volume} {13}},\ \bibinfo {pages} {449} (\bibinfo {year}
  {2019})}\BibitemShut {NoStop}%
\bibitem [{\citenamefont {Egan}\ \emph {et~al.}(2015)\citenamefont {Egan},
  \citenamefont {Jacques}, \citenamefont {Wilson}, \citenamefont {Veale},
  \citenamefont {Seller}, \citenamefont {Beale}, \citenamefont {Pattrick},
  \citenamefont {Withers},\ and\ \citenamefont {Cernik}}]{Egan2015}%
  \BibitemOpen
  \bibfield  {author} {\bibinfo {author} {\bibfnamefont {C.~K.}\ \bibnamefont
  {Egan}}, \bibinfo {author} {\bibfnamefont {S.~D.~M.}\ \bibnamefont
  {Jacques}}, \bibinfo {author} {\bibfnamefont {M.~D.}\ \bibnamefont {Wilson}},
  \bibinfo {author} {\bibfnamefont {M.~C.}\ \bibnamefont {Veale}}, \bibinfo
  {author} {\bibfnamefont {P.}~\bibnamefont {Seller}}, \bibinfo {author}
  {\bibfnamefont {A.~M.}\ \bibnamefont {Beale}}, \bibinfo {author}
  {\bibfnamefont {R.~A.~D.}\ \bibnamefont {Pattrick}}, \bibinfo {author}
  {\bibfnamefont {P.~J.}\ \bibnamefont {Withers}}, \ and\ \bibinfo {author}
  {\bibfnamefont {R.~J.}\ \bibnamefont {Cernik}},\ }\href {\doibase
  10.1038/srep15979} {\bibfield  {journal} {\bibinfo  {journal} {Scientific
  Reports}\ }\textbf {\bibinfo {volume} {5}},\ \bibinfo {pages} {15979}
  (\bibinfo {year} {2015})}\BibitemShut {NoStop}%
\end{thebibliography}%
%

\end{document}